\documentclass[journal,11pt,final,twocolumn,letterpaper]{IEEEtran}
\usepackage{url}
\usepackage[normalem]{ulem}
\usepackage[user]{zref}
\usepackage{graphicx}
\usepackage{subfigure}
\usepackage{amsmath} 
\usepackage{pstricks}
\usepackage{tikz}
\usepackage{pgfplots}
\pgfplotsset{compat=newest}
\usetikzlibrary{arrows,decorations.markings,decorations.pathmorphing}
\usetikzlibrary{patterns}

\renewcommand{\exp}[1]{\textrm{e}^{#1}}

\begin{document}
\title{Dynamic Routing for Flying Ad Hoc Networks}
 \author{
\IEEEauthorblockN{ Stefano Rosati,~\IEEEmembership{Member,~IEEE}, Karol Kru\.zelecki,~\IEEEmembership{Member,~IEEE},   Gr\'egoire Heitz, Dario Floreano,~\IEEEmembership{Senior Member,~IEEE}, and Bixio Rimoldi,~\IEEEmembership{Fellow,~IEEE} } \\
\thanks{
S. Rosati, K. Kru\.zelecki, and B. Rimoldi are with the Mobile Communications Laboratory (LCM), Swiss Federal Institute of Technology (EPFL), Lausanne,  Switzerland. 
G. Heitz and D. Floreano are with Laboratory of Intelligent Systems (LIS), EPFL, Lausanne,  Switzerland. 
Email: \{stefano.rosati, karol.kruzelecki, gregoire.heitz, dario.floreano, bixio.rimoldi\}@epfl.ch
}
}

\maketitle

\IEEEpeerreviewmaketitle

\begin{abstract}
This paper reports experimental results on self-organizing wireless networks carried by small flying robots.
Flying ad hoc networks (FANETs) composed of small unmanned aerial vehicles (UAVs) are flexible, inexpensive  and fast to deploy.
This makes them a very attractive technology for many civilian and military applications.
Due to the high mobility of the nodes, maintaining a communication link between the UAVs is a challenging task.
The topology of these networks is more dynamic than that of typical mobile ad hoc networks (MANETs) and of typical vehicle ad hoc networks (VANETs). 
As a consequence, the existing routing protocols designed for MANETs partly fail in tracking network topology changes.

In this work, we compare two different routing algorithms for ad hoc networks: optimized link-state routing (OLSR), and predictive-OLSR (P-OLSR).  
The latter is an OLSR extension that we designed for FANETs; it takes advantage of the GPS information available on board.  
To the best of our knowledge, P-OLSR is currently the only FANET-specific routing technique that has an available Linux implementation. 
We present results obtained by both Media Access Control (MAC) layer emulations and real-world experiments.
In the experiments, we used a testbed composed of two autonomous fixed-wing UAVs and a node on the ground.
Our experiments evaluate the link performance and the communication range, as well as the routing performance. 

Our emulation and experimental results show that P-OLSR significantly outperforms OLSR in routing in the presence of frequent  network topology changes.
\end{abstract}

%\pagebreak

\section{Introduction}

In the case of a calamitous event, when ordinary communication infrastructure is  out of service or simply not available, a  group of small flying robots can provide a rapidly deployable and self-managed ad hoc Wi-Fi network to connect and coordinate rescue teams on the ground. 
Networks of small unmanned aerial vehicles\footnote{Small UAVs are also knows as  micro-air vehicles (MAVs)} (UAVs) can also be employed for wildfire monitoring \cite{bib:Wildfire_monitoring},  border surveillance \cite{bib:borderpatrol}, and for extending ad hoc networks on the ground \cite{bib:other_networking_techiniques_Rubin,bib:other_networking_techiniques_Freitas,ieee5700245}.

The recent technological progress in electronics and communication systems, especially due to the the wide-spread availability of low-cost micro-embedded computers and Wi-Fi radio interfaces, 
has paved the way for the creation of inexpensive flying ad hoc networks (FANETs), %  where all, or part of the nodes, are carried by small UAVs. 
however, a challenging networking problem arises. 

As pointed out in  \cite{bib:FANETsurvey}, FANETs are a special case of mobile ad hoc networks (MANETs) characterized by a high degree of mobility.
In a FANET, the topology of the network can change more frequently than in a typical  MANET or vehicle ad hoc network  (VANET).
As a consequence, the network routing becomes a crucial task \cite{bib:routingUAVSahingoz,ieee4770039}. 
The network routing algorithms, which have been  designed for MANETs, such as  BABEL  \cite{bib:rfc-babel} or the optimized link-state routing (OLSR) protocol \cite{bib:rfc-olsr,bib:rfc-olsr2}, fail to follow the evolution of the  network topology.
It is possible to  bypass this problem by considering star networks with static routing \cite{bib:Frew2008}. However, star architectures restrict the operative  area of groups of UAVs, because the nodes cannot fly out of the communication range of the control center.
In this paper, we focus on partially-connected mesh ad hoc networks that enable the UAVs to use multi-hop communication to extend the operative area.
In this case, the problem of highly dynamic routing must be faced.

A few methods for overcoming the  problem caused by a rapidly changing network topology have recently been proposed:
Guo at al. \cite{ieee6214304} present a UAV-aided cross-layer routing protocol (UCLR) that aims at improving the routing performance of a ground MANET network with aid from one UAV. 
In \cite{ieee5461502_2}, the authors propose the use of directional antennas and two cross-layer schemes, named Intelligent Media Access Control (MAC), and Directional-OLSR.
The latter is an extension of OLSR.
The choice of the route is based on flight information (such as attitude variations, pitch, roll and yaw). The authors report only OPNET simulation results. 
Benzaid et al \cite{ieee1045725,ieee1207131} present an OLSR extension denoted Fast-OLSR that aims at meeting  the need for highly dynamic routing in MANETs composed of \emph{fast-moving} and \emph{slow-moving} nodes.
This extension increases the rate of the \emph{Hello} messages only for the nodes that move faster than a given speed. 
If the \emph{fast-moving} nodes are a small percentage of the network's nodes, the additional overhead is limited.
Otherwise, if the network is composed mainly of UAVs, the overhead grows significantly.
In \cite{bib:GC13}, we  present an extension to the OLSR protocol named predictive-OLSR (P-OLSR). % 
The key idea of this extension is to use GPS information available on board and to weigh the expected transmission count (ETX) metric by a factor that takes into account the  direction and the relative speed between the UAVs.

In this paper, we present field experiments that compare P-OLSR against OLSR. %, provide more insight into the algorithm.
The experiments involved two UAVs and a ground station. 
The carriers were fixed-wing autonomous planes called \emph{eBees} and developed by SenseFly \cite{bib:sensefly}. Each plane carried an embedded computer-on-module and an 802.11n radio interface. 
The field-test results show that P-OLSR can follow rapid topology changes and provide a reliable multi-hop communication in situations where OLSR mostly fails.
In order to assess the  behavior of P-OLSR in larger networks, we carried out MAC-layer emulations considering a network composed of 19 UAVs.
To the best of our knowledge, P-OLSR is currently the only FANET-specific routing technique that has an available Linux implementation.
The open-source software of the P-OLSR daemon can be downloaded from \cite{smavnet_website_download}.

The rest of the paper is organized as follows. 
In Section \ref{Polsdr}, we highlight the differences between OLSR and P-OLSR. 
In Section \ref{implementation}, we specify the modified structure of the {\it Hello} messages and describe the implementation of the P-OLSR daemon.
In Section \ref{test}, we describe the testbed and in Section \ref{experiments} we describe the experiments and present the results. 
In Section \ref{emulation}, we present the MAC-layer emulation we used to assess the P-OLSR performance in larger networks. 
Finally, we draw our conclusions in Section \ref{conclusions}.

\section{ Routing for Flying Ad Hoc Networks}
\label{Polsdr}

In \cite{bib:Frew2008}, Frew and Brown analyze the networking for systems of small UAVs.
They characterize four different network architectures: \emph{direct-link}, \emph{satellite}, \emph{cellular}, and \emph{ad hoc} (also called  \emph{mesh networking}). 
The \emph{direct-link} and the \emph{satellite} architecture are star networks where all the UAVs are  either directly connected 
to the ground control or to a satellite connected to the ground control.
This is a simple network architecture: it does not require dynamic routing, because all the nodes are directly connected with the control center. 
The nodes require, however,  long-range (terrestrial or satellite) links, hence they are not suitable for small UAVs.
Furthermore, UAV-to-UAV communication is inefficiently routed through the control center, even if the nodes operate the same area.
This might cause traffic congestion in the control center, which is also a weakness of the system in case of attack.  

In the \emph{cellular} architecture, the UAVs are connected to a cellular system with many base stations scattered on the ground. 
UAVs can do a handover between different base stations during the flight. 
This architecture does not need a single vulnerable control center.
However, the operating area of the UAVs is limited by the cellular network extension.
In the case of a catastrophic event, the UAV system can be deployed only if the cellular network is present and functioning in the area.

In the \emph{ad hoc} architecture, every node can act as a router. These networks are also know as FANETs: 
they have no central infrastructure, therefore,  they are very robust against isolated attacks or node failures.
Moreover, as these networks do not rely on any external support they can be rapidly deployed anywhere.
These characteristics, on one hand, make FANETs the most suitable solution for many applications, 
but on the other hand, they raise a challenging networking problem.

In fact, due to the rapid and erratic movement of the UAVs, the topology of a FANET can vary rapidly and the nodes must react by automatically updating their routing tables.
Therefore, in  a FANET it is crucial to employ a fast and reactive routing procedure. 
In \cite{bib:GC13} we show that some of the most popular routing algorithms for  MANETs, such as  OLSR and BABEL, fail to track the fast topology changes of a FANET.
Similar conclusions are also drawn in \cite{bib:routingUAVSahingoz}. 
For this reason, in \cite{bib:GC13}, we introduce  P-OLSR. 
To predict how the quality of the wireless links between the nodes is likely to evolve, P-OLSR exploits the GPS information, which is typically available from the UAV's autopilot.
For the sake of completeness, in this section we briefly report the definition of ETX, as well as, the key concepts behind P-OLSR.

\subsection{Link-Quality Estimation}

OLSR is currently one of the most popular proactive routing algorithms for ad hoc networks. It is based on the link-state routing protocol. 
The original OLSR design does not consider the quality of the wireless link. The route selection is based on the hop count metric, which  is inadequate for mobile wireless networks. 
However, by using the ETX metric \cite{bib:olsrdLQ}, the OLSR \emph{link-quality} extension enables us to take into account the quality of the wireless links. 
The ETX metric was introduced in \cite{bib:Couto03}, and it is defined as
\begin{equation}
 \mathrm{ETX}(\mathcal{R})= \sum_{\eta \in \mathcal{R}} \mathrm{ETX}(\eta) \, = \sum_{\eta \in \mathcal{R}}  \frac{1}{\phi(\eta)  \rho(\eta)} \, , 
\end{equation}
where, $\mathcal{R}$ is a route between two nodes of the network, and $\eta$ is a hop of the route $\mathcal{R}$.
$\phi(\eta)$ is the forward receiving ratio, i. e., the probability that a packet sent through the hop $\eta$ is successfully received.
$\rho(\eta)$ is the reverse receiving ratio, i.e., the probability that the corresponding ACK packet is successfully received.
In other words, ETX estimates the expected number of transmissions (including re-transmissions) necessary to deliver a packet from the source to its final destination.
Then OLSR selects the route that has the smallest ETX, which is not necessarily the one with the least number of hops. 
If all the hops forming $\mathcal{R}$ are errorless (i.e., $\phi(\eta)=\rho(\eta)=1$) the $\mathrm{ETX}(\mathcal{R})$ is equal to the number of hops of $\mathcal{R}$.

The receiving ratios are typically estimated by link-probe messages.
The OLSR link-quality extension uses the control messages named \emph{Hello} messages as a link-probe. $\phi$ is computed by means of an exponential moving average, as follows,
\begin{equation}
 \begin{cases}
 \phi_{l} =  \alpha h_{l} + (1-\alpha) \phi_{l-1} \\ 
 \phi_{0}=0
\end{cases}
\! \! \! ,\quad 0\leq\alpha\leq1 \, ,
 \end{equation}
 where
 \begin{equation}
 h_{l} = \begin{cases}
1 & \mbox{if the  $l$-th \emph{Hello} message is received} \\
0 & \mbox{otherwise}  
\end{cases}\, ;
\end{equation}
and $\alpha$ is an OLSR parameter, named \emph{link-quality aging}, that  drives the trade-off between the accuracy and responsiveness of the receiving ratio estimation.
On one hand,  with a greater $\alpha$, the receiving ratios will be averaged for a longer time, thus yielding a more stable and reliable estimation.
On the other hand,  with a lower  $\alpha$, the system will react faster. 
Another important OLSR parameter is the  \emph{Hello Interval} (HI) that indicates how frequently \emph{Hello} messages are broadcasted.

\subsection{Speed-Weighted ETX}

In a FANET the nodes move rapidly, for example, the cruising speed of our flying robots is around 12 meters per second. 
The network topology changes rapidly, and a wireless link between two UAVs can break suddenly.
In these networks, the ETX metric might be inadequate, because it is not reactive enough to follow the link variations.
Due to the delay introduced by the exponential moving average, a node notices that a certain wireless link has broken with a non-negligible delay.
Therefore, for a significant amount of time, it will continue routing packets on a link that is actually broken.

To solve this problem, we modify the $\mathrm{ETX}$ metric to take into account the position and the direction of the UAV, with respect to its neighbors.
The $\mathrm{ETX}(\eta)$ metric of the hop $\eta$ between the nodes $i$ and $j$ is weighed by a factor that accounts for the  relative speed between $i$ and $j$ as follows:
\begin{equation}\label{etx2}
 \mathrm{ETX}(\eta) = \frac{\exp{v_{\ell}^{i,j}  \beta}}{\phi(\eta)  \rho(\eta)} \, ,
\end{equation}%
where $v_{\ell}^{i,j}$ is the relative speed between nodes $i$ and $j$, and  $\beta$ is a non-negative parameter.

If the nodes $i$ and $j$  move closer to each other, the relative speed is negative, thus the ETX will be weighted by a factor smaller than 1. 
Otherwise, if the nodes $i$ and $j$ move apart from each other, the relative speed is positive, thus the ETX will be weighted by a factor greater than 1. 
In other words, a hop between two nodes that move closer to each other is preferred rather than a hop between two nodes that move apart, even if they have the same values of $\phi$ and $\rho$.

In order to compute the speed-weighted  ETX, we  assume that every node knows the position of its neighbors. 
As illustrated in the Appendix \ref{implementation}, to  distribute the GPS coordinates across the network we add a field to the \emph{Hello} messages.

The instantaneous relative velocity between $i$ and $j$ at time $t_i$ is computed as
\begin{equation}\label{ispeed}
 \tilde{v}_{\ell}^{i,j}= \frac{ d_{\ell}^{i,j} - d_{\ell-1}^{i,j} }{t_{\ell} - t_{\ell-1}} \, ,
\end{equation} 
where, $t_{\ell}$ and $t_{\ell-1}$ are,  the arrival time of the last and second to last \emph{Hello} message.  $d_{\ell}^{i,j}$  and  $d_{\ell-1}^{i,j}$ are the corresponding distances between the nodes $i$ and $j$. 
As the GPS positions are subject to errors, and gusts of wind can perturb the  motion of the UAVs,
it is preferable to average the instantaneous speed using  a exponential moving average as follows:

\begin{equation}
 \begin{cases}\label{rspeed}
   v_{\ell}^{i,j} =  \gamma \tilde{v}_{\ell}^{i,j}  + (1-\gamma) v_{\ell-1}^{i,j} \\
   v_{0}^{i,j} = 0  
 \end{cases}
 ,\quad 0\leq\gamma\leq1 \, ,
 \end{equation}
where $\gamma$ is a P-OLSR parameter.
Acting on the P-OLSR parameter $\beta$ and $\gamma$, we can optimize the routing selection to the cruising speed of the UAVs, and to the chosen $\mathrm{HI}$.

Fixed-wing UAVs require forward motion for flying. They also require a minimum air-speed and turning radius. 
Therefore the direction of the UAV is a good indicator for predicting its position in the near future, and then to foresee how the link quality is likely to evolve.

\section{Implementation Details}
\label{implementation}

In order to implement the P-OLSR protocol, we forked an open-source implementation of OLSR called OLSRd \cite{bib:olsrd}. 
In the modified version, the \emph{Hello} messages are augmented to contain position information. Thus, every node knows its neighbors' positions and can compute the corresponding ETX according to \eqref{etx2}.

\subsection{OLSRd with Link-Quality Extension}

OLSRd uses link-quality sensing and ETX metrics through the so-called \emph{link-quality} extension \cite{bib:olsrdLQ}.
It replaces the hysteresis mechanism of the OLSR protocol with link-quality sensing algorithms that are intended to be used with ETX-based metrics.
To do so, the \emph{link-quality} extension uses the OLSR \emph{Hello} messages to probe link quality and to advertise link-specific quality information, (i.e. receiving ratios, $\phi$ and $\rho$), in addition to detecting and advertising neighbors.
Likewise, it includes the link-quality information also in OLSR Topology Control (TC) messages that are to be distributed  to the whole network.
Clearly the modified messages are not RFC-compliant anymore because they include new fields for link-quality information. 
Therefore, all the nodes in the network have to use the link-quality extension.

\subsection{P-OLSRd Implementation}

To implement P-OLSR, we have to share the coordinates (i.e. longitude, latitude, and altitude) of each node with its neighbors. 
This is done through the {\it Hello} messages.
Subsequently, each node uses its neighbors' coordinates to compute the corresponding relative speeds, and share them across the whole network via both {\it Hello} and TC messages.

\begin{figure}
\centering
\scalebox{0.6} % Change this value to rescale the drawing.
 {
\includegraphics[width=10cm]{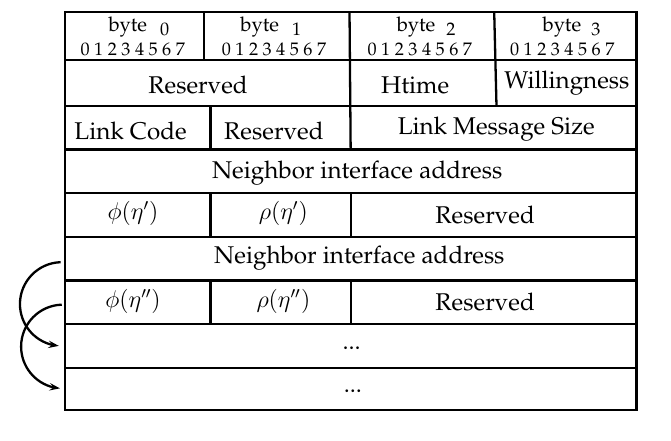}
 }
\caption{Format of the original (i.e. OLSRd) \emph{Hello} message. }
\label{fig:Hello1}
\vspace{-0.3cm}
\end{figure}

Fig. \ref{fig:Hello1} depicts the structure of the original {\it Hello} message in OLSRd. The first block of 8 bytes carries information about the node itself. Note that in this block there are 3 reserved bytes that are not used by the OLSR daemon and are filled with zeros. 
Then, a block of 8 bytes is appended for each neighbor seen by the node. This block is formatted as follows:
4 bytes are for the IP address of the neighbor, 1 byte is for the forward receiving ratio\footnote{$\eta^\prime$ is the hop between the node the is producing the {\it Hello} message and its neighbor.}  $\phi(\eta^\prime)$, 1 byte is for the reverse  receiving ratio  $\rho(\eta^\prime)$, and 2 bytes that are not used, hence filled with zeros.

\begin{figure}
\centering
\scalebox{0.6} % Change this value to rescale the drawing.
{
\includegraphics[width=10cm]{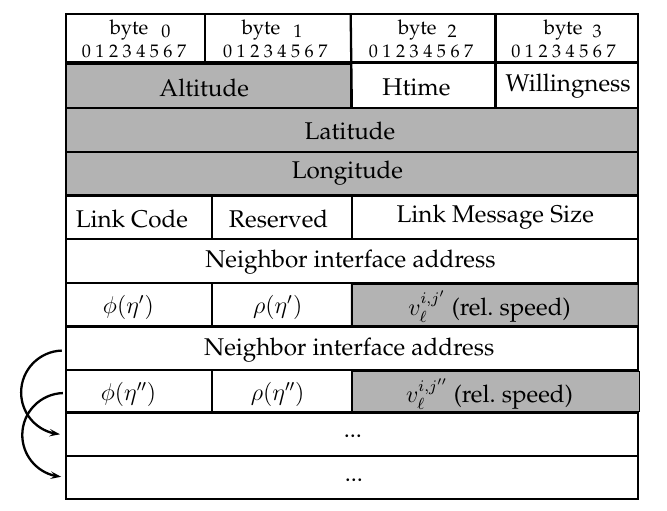}
 }
\caption{Format of the modified  (i.e. P-OLSRd) \emph{Hello} message.}
\label{fig:Hello2}
\vspace{-0.3cm}
\end{figure}

Fig. \ref{fig:Hello2} depicts the modified structure of the {\it Hello} message. We highlight in gray the fields that were added or modified. 
The first part of the message contains 16 bytes instead of 8.
It contains the latitude and the longitude, formatted as single-precision floating-point numbers that occupy 4 bytes each\footnote{In order to have at least one-meter precision, we could have formatted the latitude and longitude with  26-bit fixed-point representation each. This, however,  would have required rearranging  the latitude and longitude, and adding some padding in order to complete the byte.}.
The altitude is formatted as a 16-bit fixed-point number that replaces the 2 reserved bytes not used by the OLSR daemon. 
Similarly to the original {\it Hello} message, a block of 8 bytes is appended for each neighbor seen by the node. 
The only difference here is that we use the 2 empty bytes to communicate the averaged relative speed between the nodes. The speed is formatted as a 16-bit fixed-point number.

The size difference between the original and the modified {\it Hello} message is 8 bytes, independently of the numbers of nodes in the network. 
The {\it Hello} message is encapsulated into a UDP datagram that, in turn, is encapsulated in an IP packet and then into a 802.11 frame.
For medium and  large  networks, the additional 8 bytes constitute a negligible overload compared to the total size of the frame.

\begin{figure}
\centering
\scalebox{0.6} % Change this value to rescale the drawing.
{
\includegraphics[width=10cm]{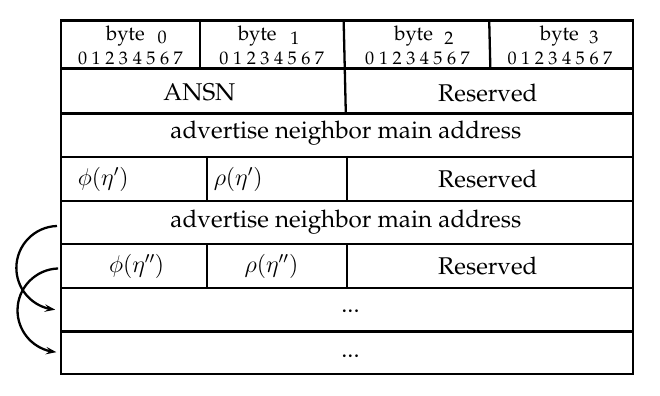}
 }
\caption{Format of the original (i.e. OLSRd) \emph{Topology Control} message. ANSN stands for Advertised Neighbor Sequence Number. }
\label{fig:TC1}
\vspace{-0.3cm}
\end{figure}

\begin{figure}
\centering
\scalebox{0.6} % Change this value to rescale the drawing.
{
\includegraphics[width=10cm]{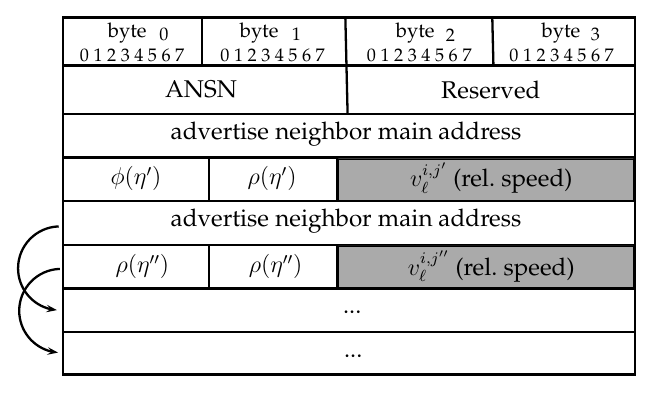}
}
\caption{Format of the modified (i.e. P-OLSRd) \emph{Topology Control} message.}
\label{fig:TC2}
\vspace{-0.3cm}
\end{figure}

Figs. \ref{fig:TC1} and \ref{fig:TC2} illustrate the structure of the TC message for OLSRd and P-OLSRd respectively. They differ only by one field: P-OLSRd exploits the 2 reserved bytes to convey the averaged relative speed formatted as a 16-bit fixed-point number. The original and the modified TC message have the same size.

\section{UAV Testbed}\label{test}

We can distinguish two main types of small UAVs:  rotary-blade and fixed-wing.
Rotary-blade UAVs, which are also know as mini-copters or multi-copters, fly using the lift force generated by one or more blades, like helicopters.
Typically, rotary-blade UAVs are composed of four blades, in this case they are know as quadri-copters.
They are able to takeoff and land vertically, fly in every direction and maintain a fixed position in the air.

Fixed-wing UAVs fly  by exploiting  the lift force generated by the forward speed of the vehicle in the air, hence, forward motion is required for flying.
They have a minimum speed and  turning radius and can fly at higher speeds and with a higher energy efficiency compared to rotary-blade UAVs. 
Therefore, they can cover a greater area, but the network topology is likely to change very rapidly.
This makes the ad hoc networking of fixed-wing UAVs more challenging and, at the same time,  more interesting.

\begin{figure}[!htp]
	\centering
	 \subfigure[SenseFly \emph{eBee} with HD camera, Linksys AE3000 USB dongle, and Gumstix computer-on-module. ]
%	 {	\includegraphics[width=0.9\columnwidth]{ebee_inkscape.eps}  \label{fig:ebee1}}
	 {	\includegraphics[width=0.9\columnwidth]{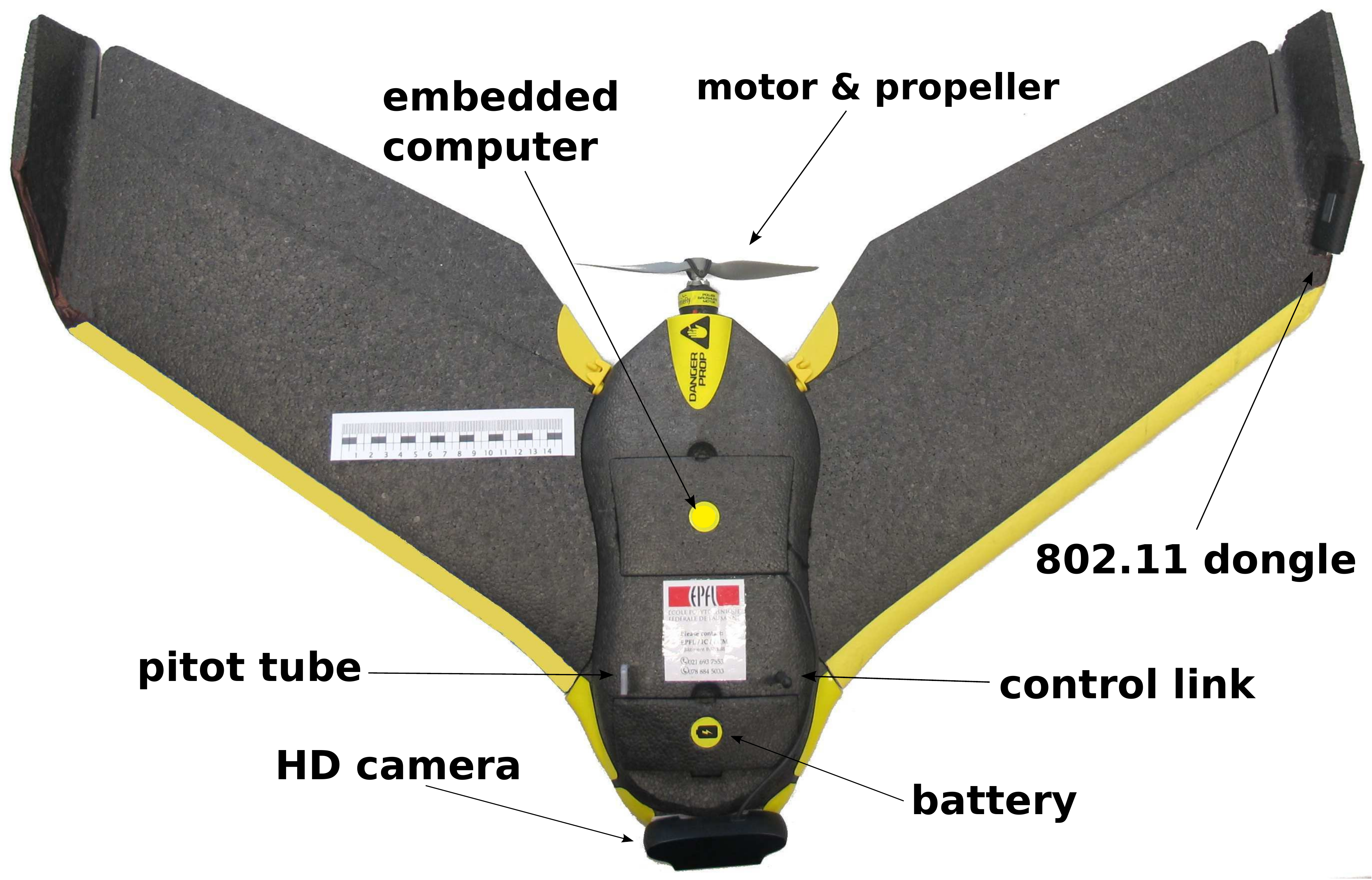}  \label{fig:ebee1}}
 \subfigure[Computer and expansion board in the rear compartment of the UAV.]	
% {      \includegraphics[width=0.9\columnwidth]{comp_inkscape.eps} \label{fig:ebee2} }
 {      \includegraphics[width=0.9\columnwidth]{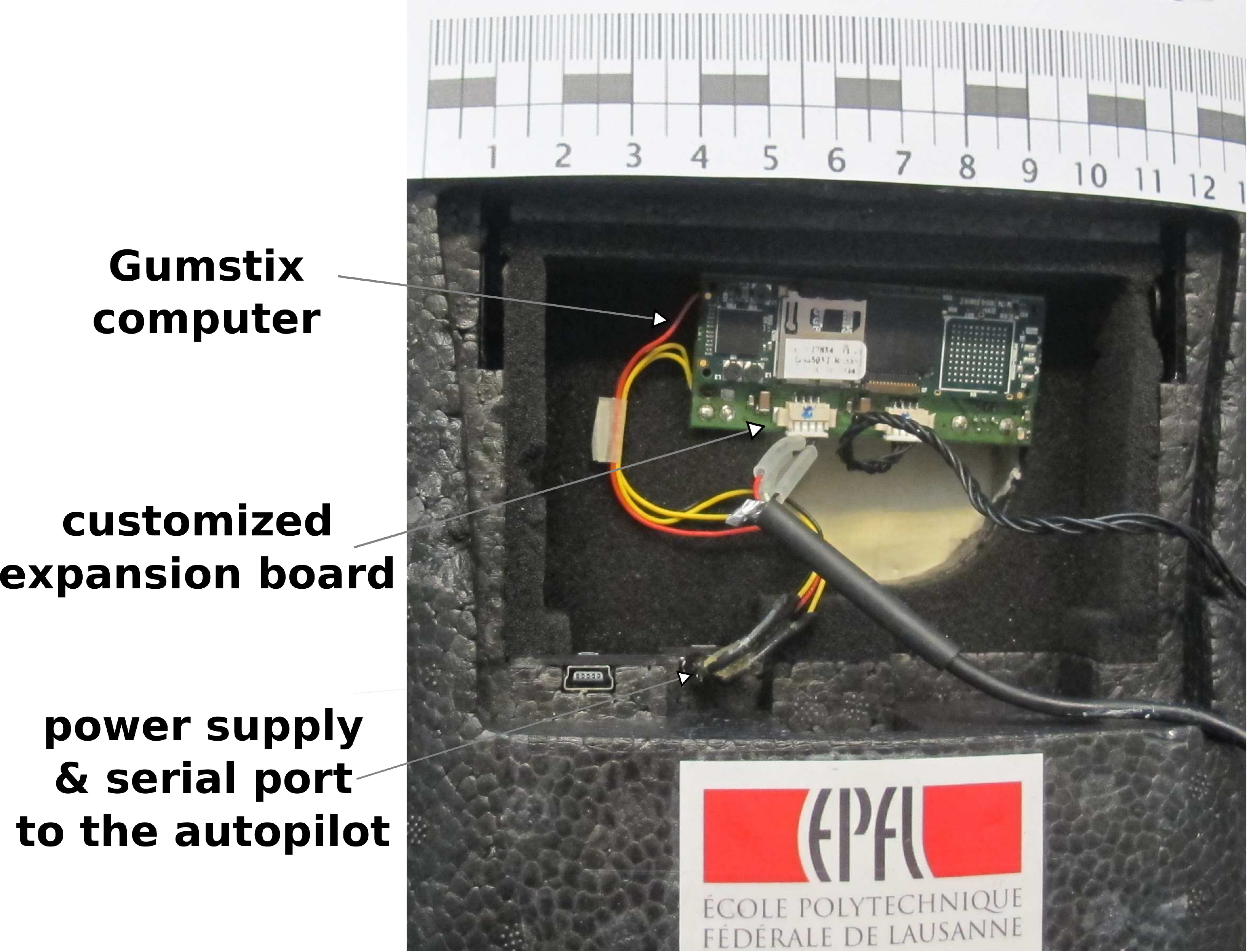} \label{fig:ebee2} }
	\caption{ UAV platform used in the experiments. } % for one- and two-hops scenarios
	\label{fig:ebee}
\end{figure}

In this work, we address fixed-wing UAVs instead of those with rotary-blades.
For the experiments we use two fixed-wing UAVs, named \emph{eBee}, developed by SenseFly \cite{bib:sensefly}. 
 The vehicle's body is made of expanded polypropylene (EPP), and it has a single rear-mounted propeller powered by an electric motor. The \emph{eBee} platform is illustrated in Fig. \ref{fig:ebee1}.
It has an integrated autopilot capable of flying with winds up to 12 m/s, at a cruising speed of about 57 km/h, with an autonomy of 45 minutes. 
In case of emergency, they can be remotely controlled up to a distance of 3 km via a Microhard Systems Nano n2420 \cite{bib:microhard-n2420} link connection, 
and this low data-rate  control link is only used for controlling the plane. The data produced by the UAV is instead carried by the 802.11n link. 
Due to its small dimensions (the wingspan is 96 centimeters)  and weight (under 630 g), flying \emph{eBees} are not considered to be dangerous. In some countries (e.g., Switzerland) they can be used without specific authorization.

On each plane, we mount a Gumstix Overo Tide \cite{bib:overoTide}, an ARM-based computer-on-module produced by Gumstix Inc. 
The computer runs a customized  Linux distribution (kernel version 3.5.0), and it is connected to a compatible expansion board that we designed and produced. 
The expansion board incorporates a USB 2.0 hub with four powered ports and a serial port connected to the autopilot. 
Using this serial port, the Gumstix can fetch the current GPS data, as well as other flight parameters. 
The computer is also connected via USB to a HD camera and  to a  Wi-Fi radio interface. 
The embedded computer setup is shown in Fig. \ref{fig:ebee2}.

The Wi-Fi radio interface is the  Linksys AE3000 USB dongle, using 802.11n. 
Despite the small dimensions, they support a MIMO system with three antennas. 
During the experiments, we enabled 802.11n space-time block coding (STBC) to exploit channel diversity.
The transmission bandwidth was set to 20 Mhz in the 5 Ghz unlicensed band.
We used QPSK modulation, code rate 1/2, and  one spatial stream (MODCOD 1). 

\section{Experiments}\label{experiments}

In this section, we describe the two experiments that we carried out on the EPFL campus. 
The aim of the first experiment was to characterize the end-to-end link performance between an UAV and a node on the ground.
Specifically, we  determined the communication range in which the two nodes can communicate with a tolerable amount of lost packets.
In the second experiment, we compared the OLSR protocol with P-OLSR protocol.

\subsection{Link Performance Assessment}\label{link}

In this experiment, we had a  2-node network composed of a node on the ground and a flying UAV.
The UAV flew between two checkpoints positioned 450 meters apart. 
For both checkpoints, we set a turning radius of 30 meters, and the height-above-ground was 75 meters. 
The node on the ground was located at the center of one of the checkpoints.
Using \emph{iperf}, we took one measurement of the link quality per second. 
Every second, the  UAV sent 85 UDP datagrams (totaling 1 Mbit) to the node on the ground. 
All the datagrams received with a delay greater than 5 seconds were considered lost. 
We also counted as correctly received the datagrams that arrived out of order (provided that their delay was smaller than 5 seconds). 
This is what can be tolerated by a video streaming, where the video is played with a 5-second delay.
Every second, we computed the datagram loss rate (DLR), which is the ratio between the datagram lost and their total number.

\begin{figure}
\centering
% \scalebox{1.0}{ % Change this value to rescale the drawing.
% \begin{tikzpicture}
% \begin{axis}[xmin=0,xmax=500,ymin=0,ymax=1,xlabel={distance [m]},ylabel={DLR},legend pos=north west]
%     \addplot[no markers,color=red,densely dashed] table[x=xreg,y=yreg] {data/xreg.dat} ;
% \addplot[only marks,mark=+] table[x=distance,y=per_distance] {data/per_distance.dat} ;
% \legend{{\small regression function},{ \small experimental DLR}}
% \end{axis}
% \end{tikzpicture} }
 \includegraphics[width=0.9\columnwidth]{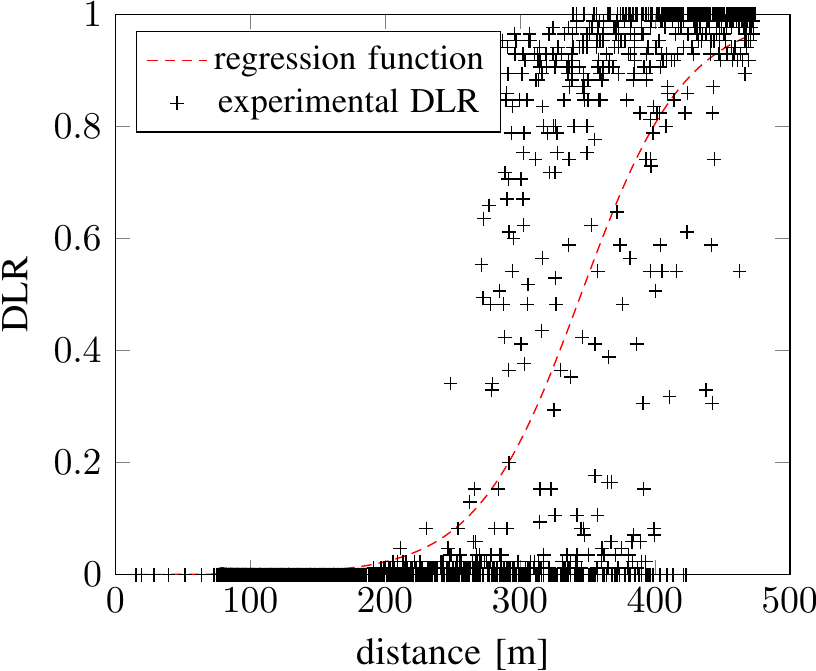}
  \caption{ Experimental datagram loss rate vs. distance. The black crosses are the DLR measured values, and the dashed red line is the corresponding non-linear least-square regression function. }
  \label{fig:dlr_link}
\end{figure}

The results are reported in Fig. \ref{fig:dlr_link}.
The black crosses represent the DLR measured at a given distance. 
The  dashed line represents the corresponding non-linear least-square regression function.
The function that we use  as a model is
\begin{equation}
\frac{1}{1+\exp{- (p_1+p_2 d)}} \, , 
\end{equation} 
where $d$ is the distance in meters and $p_1$ and $p_2$ are the coefficients to be estimated.
Using a non-linear fitting method, we compute the value of $p_1$ and $p_2$, thus yielding the best fitting in the least square sense.
They are  $p_1=8.9$ and $p_2=0.025$.

\begin{figure}
 \centering
% \scalebox{1.0}{ % Change this value to rescale the drawing.
% \begin{tikzpicture}
% \pgfplotsset{every axis/.style={ymin=0}}
% \begin{axis}[ybar, bar width=6pt,xmin=0,xmax=500, axis y line*=right, axis x line=none, ylabel={Number of measurements},legend pos=north west]%
%   \addplot[blue,no markers]  table[x=distance_vector_out,y=histogram_distances] {data/mean_per_distance.dat} ;
% \legend{{\small Number of measurements}}
% \end{axis}
% \begin{axis}[xmin=0,xmax=500,ymin=0,ymax=1,xlabel={distance [m]},ylabel={DLR},legend style={at={(0.1,0.8)},anchor=west} ]
%   \addplot[mark=diamond,color=green!50!black] table[x=distance_vector_out,y=mean_per_distance] {data/mean_per_distance.dat} ;
%   \legend{{\small Average DLR}}
% \end{axis}
% \end{tikzpicture} }
 \includegraphics[width=0.9\columnwidth]{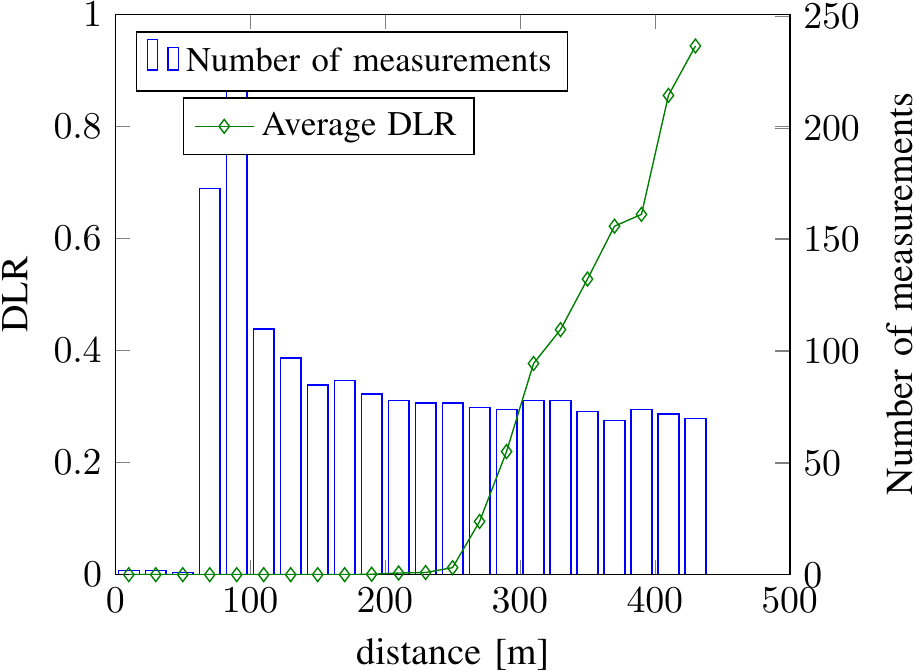}
 \caption{Average datagram loss rate vs. distance. The blue bars represent the number of measurements per distance interval.  The green solid line is the average DLR.}
 \label{fig:avg_dlr_link}
 
\end{figure}

In Fig. \ref{fig:avg_dlr_link}, we quantize the distance between transmitter and receiver with a bin width of 20 meters.
For each bin we compute the average DLR.
The resulting curve is plotted in green. We also report the number of measurements per bin.

As we can see from these results, the connection is good when the distance is shorter than 250 meters. In this region, the observed DLR is always lower than 0.2.
We observe a transition from 250 meters to 300, where the DLR can be as high as 1,
but on average it is lower than 0.3. 
When the distance is greater than 300 meters, the connection degrades significantly and the  DLR is often close to 1.
The average DLR is 0.5 at 350 meters.  
We observe some sporadic cases of a good connection even when the distance is greater than 400 meters. 

\subsection{Routing Performance Assessment}

In the second experiment, we  compared the routing performance of OLSR with link-quality extension and P-OLSR.
For both OLSR and P-OLSR, we set the $\mathrm{HI}$ equal to 0.5 seconds.
This value is a good trade-off between the amount of overhead and the reactivity-speed  of the algorithm. 
Using the default $\mathrm{HI}$ value, which in OLSRd is equal to 2 seconds, the algorithm would have been too slow to pursue the topology changes. 
As P-OLSR exploits the GPS information to estimate the link-quality evolution, it could work also with a longer $\mathrm{HI}$.
Nevertheless, for the sake of fairness, we use the same value for both algorithms. 

We set the \emph{link-quality aging}, $\alpha$, equal to 0.2 for OLSR, and to  0.05 for P-OLSR. 
This parameter controls the trade-off between  the accuracy and the responsiveness of the receiving ratio estimation.
As before, as P-OLSR can predict the link-quality evolution, it can work with a smaller $\alpha$, yielding a more accurate estimation of the receiving ratio.
OLSR, on the contrary,  needs an greater aging value to increase its responsiveness, at the expense of accuracy.
The other P-OLSR parameters were  $\beta=0.2$, and  $\gamma=0.04$.

\begin{figure}
\hspace{-1.0cm}
%\scalebox{0.95} % Change this value to rescale the drawing.
\centering
 \includegraphics[width=1.0\columnwidth]{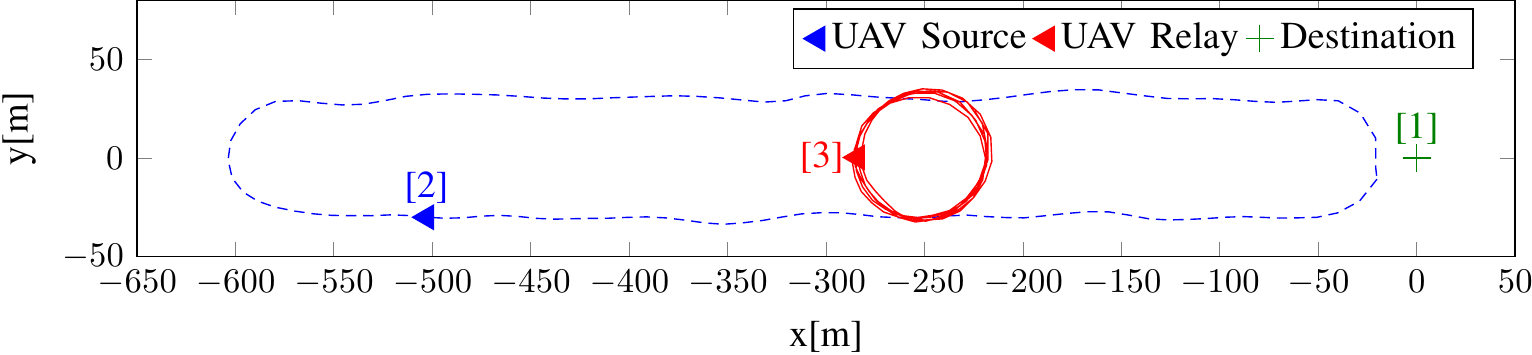}
 \caption{Trajectories of the UAVs during the network field experiments.}
 \label{fig:2relays_exp}
\end{figure}

We had a network of three nodes:
one fixed destination on the ground (node 1),  one flying UAV source (Node 2), and  one flying UAV relay (Node 3).
Node 1 was on the terrace of the BC building on the EPFL campus (latitude: $46.51843^\circ$ N;  longitude: $6.561591^\circ$ E).  
The UAV source, Node 2 flew following a straight trajectory of 600 meters west from Node 1, then it returned to the starting point. 
The UAV relay, Node 3, followed a circular trajectory of radius 30 meters and centered at 250 meters west from Node 1.
Both the UAVs flew about 75 meters above the ground, and the terrace of the BC building is approximately 10 meters above the ground. 
The actual trajectories followed by the planes are illustrated in Fig. \ref{fig:2relays_exp}.

We  performed 10 loops for each routing algorithms.
The loop-time is affected by the weather conditions, especially the wind. 
During our experiment, the eBee took on average 120 seconds to complete a loop.
The time difference between the fastest and the slowest loop was around 10 seconds.
In each loop, Node 2 changed its routing to reach Node 1,  from a direct connection to a two-hop connection, and vice versa. 
Therefore the network topology was expected to change two times during the each loop.  

\begin{figure}
\centering
\includegraphics[width=1.0\columnwidth]{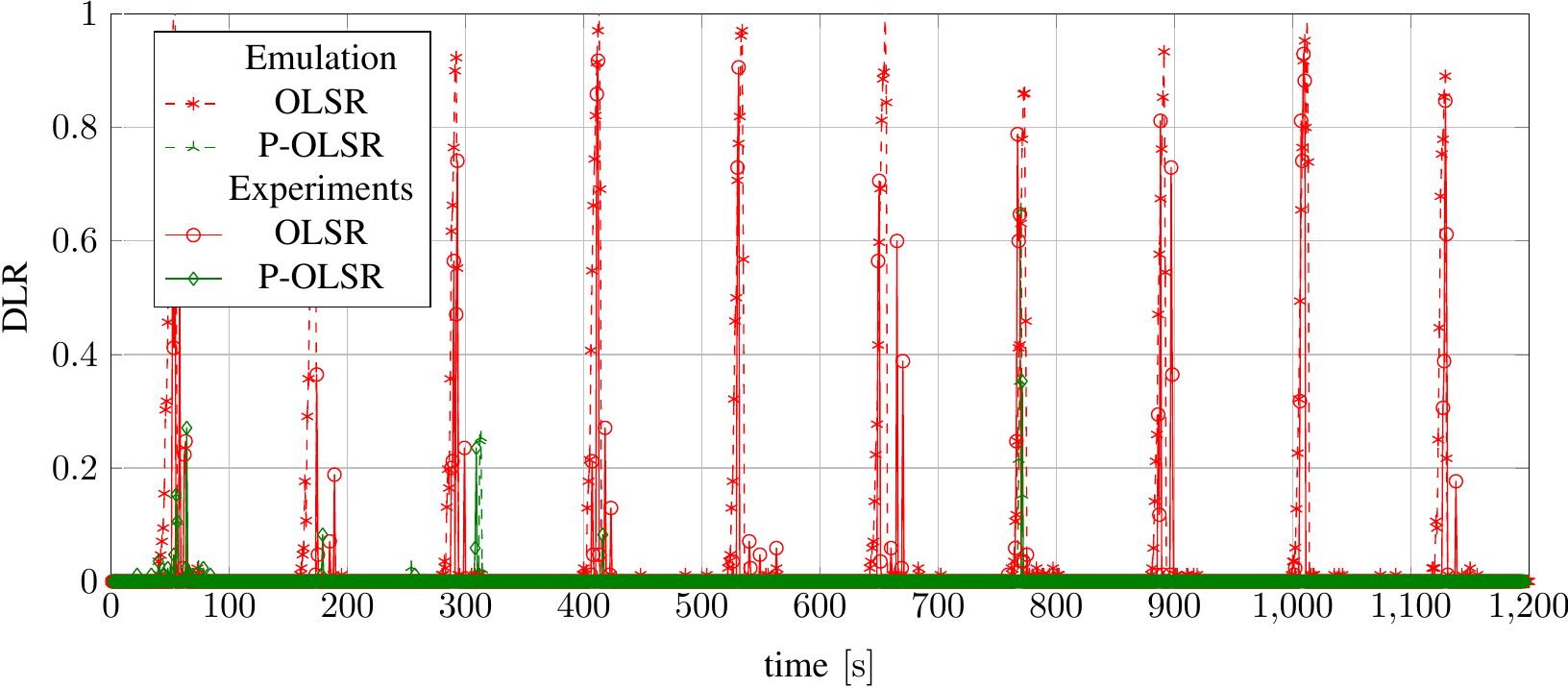}
\caption{Evolution of the average DLR. Dashed lines report the results of MAC-layer emulations, solid lines report the results of field experiments.}
 \label{fig:per_routing}
\end{figure}

Fig.  \ref{fig:per_routing} shows the evolution of the DLR during the 10 runs. 
For OLSR, we  notice some peaks of the DLR that correspond to the moments when the routing algorithm has to switch from the direct link to a two-hop.
This happens because OLSR takes several seconds to detect that a wireless direct-link is broken. 
This translates into an interruption of the service.
P-OLSR, however, reacts promptly to topology changes. As we can see from the field-test results, P-OLSR is able to predict a change in the topology and reacts  before  the previous link breaks. 
The DLR peaks that we see on the P-OLSR results are due to the fading of the wireless channels rather than to incorrect routing.

\section{MAC-layer emulations with Larger Networks}
\label{emulation}

In this section, we examine the behavior of P-OLSR operating on larger FANETs. 
When many UAVs are involved, field experiments become expensive.
For this reason, we analyze the performance of P-OLSR via a network emulation platform that integrates all the testbed aspects.

\subsection{Emulation Platform}

\begin{figure}
\centering
%\scalebox{0.80} % Change this value to rescale the drawing.
%{
%\input{emu}
\includegraphics[width=1.0\columnwidth]{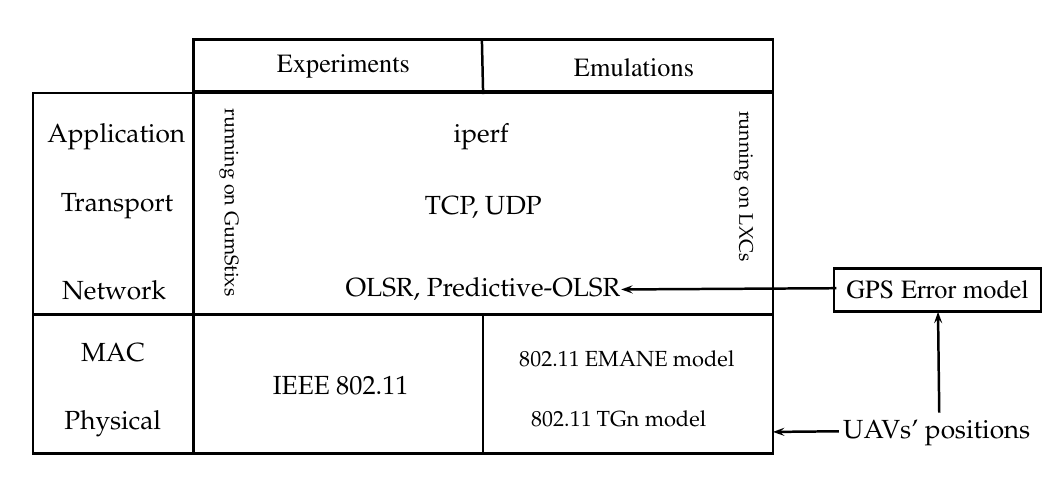}
%}
\caption{Emulation platform.}
\label{fig:emulator}
%\vspace{-0.3cm}
\end{figure}

In order to analyze the routing performance in medium/large FANETs, we developed the emulation platform illustrated in Fig. \ref{fig:emulator}.
It creates a  Linux container (LXC) for each node of the network.  The nodes are connected using a MAC-layer real-time emulator called Extendable Mobile Ad-Hoc Network Emulator (EMANE). EMANE is an open-source framework, developed primarily  by Naval Research Laboratory \cite{bib:emane}.  
The MAC and the physical layers are emulated, whereas the remaining layers use the real software implementations used by the Linux machine.
For a propagation channel model, we use the IEEE 802.11 TGn model D, defined in  \cite{bib:802.11TGn}. This model was proposed for outdoor environments with line-of-sight conditions.
EMANE imports the positions of the UAVs from log files and uses them to compute the pathloss for each link. 
These log files can be obtained from real-flight data logs, or  by a flight simulator that reproduces realistic flight conditions. 
Before passing the UAV's position to the P-OLSR daemon, we add an error to take into account imperfect GPS receivers. 
We model GPS errors following the statistical characterization of GPS error provided in \cite{GPSError}.

\subsection{Emulation Results}

We carried out two network emulation campaigns. 
In the first one, we reproduced the  conditions we had in the experiments. This was done to test the emulator.
We used the positions reported in the log files of the actual experiments.
Fig. \ref{fig:per_routing} shows emulation results (dashed lines) side-by-side with the experimental results (solid lines) for OLSR and for P-OLSR.
The emulation results match with the experimental ones.

\begin{figure}
\centering
\includegraphics[width=0.9\columnwidth]{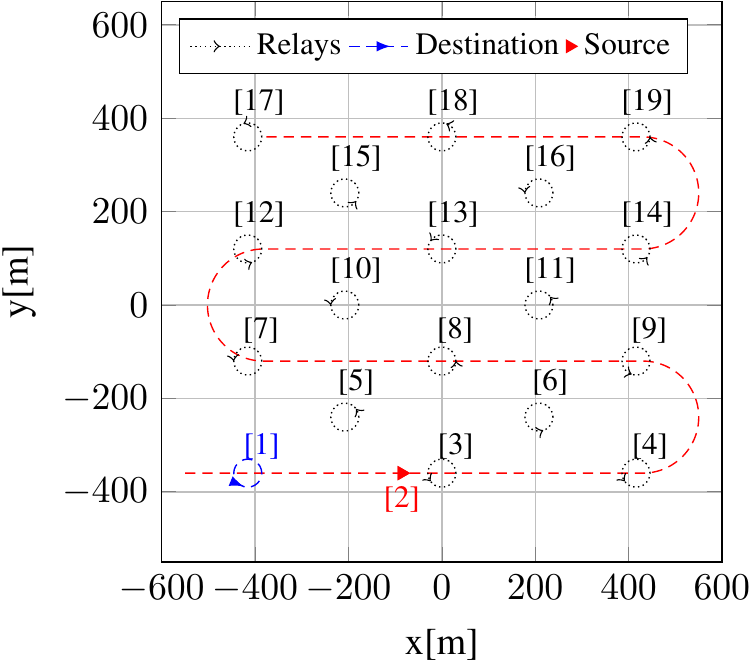}
 \caption{Emulations set-up. A network composed of 19 nodes. Node 2 is flying with a speed of 12 m/s following the trajectory markers with red dashed line.}
 \label{fig:24relays}
\end{figure}

In the second emulation campaign, we assessed the behavior of P-OLSR in larger networks and we analyzed the role of various parameters, namely
the {\it Hello interval} (HI), the link-quality aging ($\alpha$), and the P-OLSR specific parameters ($\beta$ and $\gamma$).
The network consisted of 19 moving UAVs. One UAV (Node 2) scanned a rectangular area of 1200 square meters,  by following the trajectory plotted with a dashed line in Fig. \ref{fig:24relays}. It took 380 seconds to complete the trajectory.
The other UAVs (Nodes $1, 3,\dots,19$) are uniformly spread within the area. They circulated around the respective waypoints, reproducing the behaviour of actual fixed-wing UAVs. The radius of the  circular trajectories was 30 meters, the speed was 12 m/s and the initial phase was uniformly distributed between zero and $2\pi$.
The distance between the relays was such that only the closest neighbors had a direct link. 
For example, node 10 can communicate directly only with Nodes 5, 7, 8, 12, 13, and 15.

As we did in the experiments, we sent 85 UDP datagrams per second (totaling 1 Mbit) from Node 2 to Node 1, using \emph{iperf}. 
In order to have a more compact representation of the results, we compared the network performance  in terms of average outage time.
We say that an outage occurs when the DLR becomes greater than 0.2.  
The outage duration of such an event is the length of the interval during which the DLR stays above 0.2. 
The outage time of a run is the sum of all the outage durations.
In order to average the results, we repeated the emulation 10 times for each configuration.

\begin{figure}
\centering
%\scalebox{0.95} % Change this value to rescale the drawing.
\includegraphics[width=1.0\columnwidth]{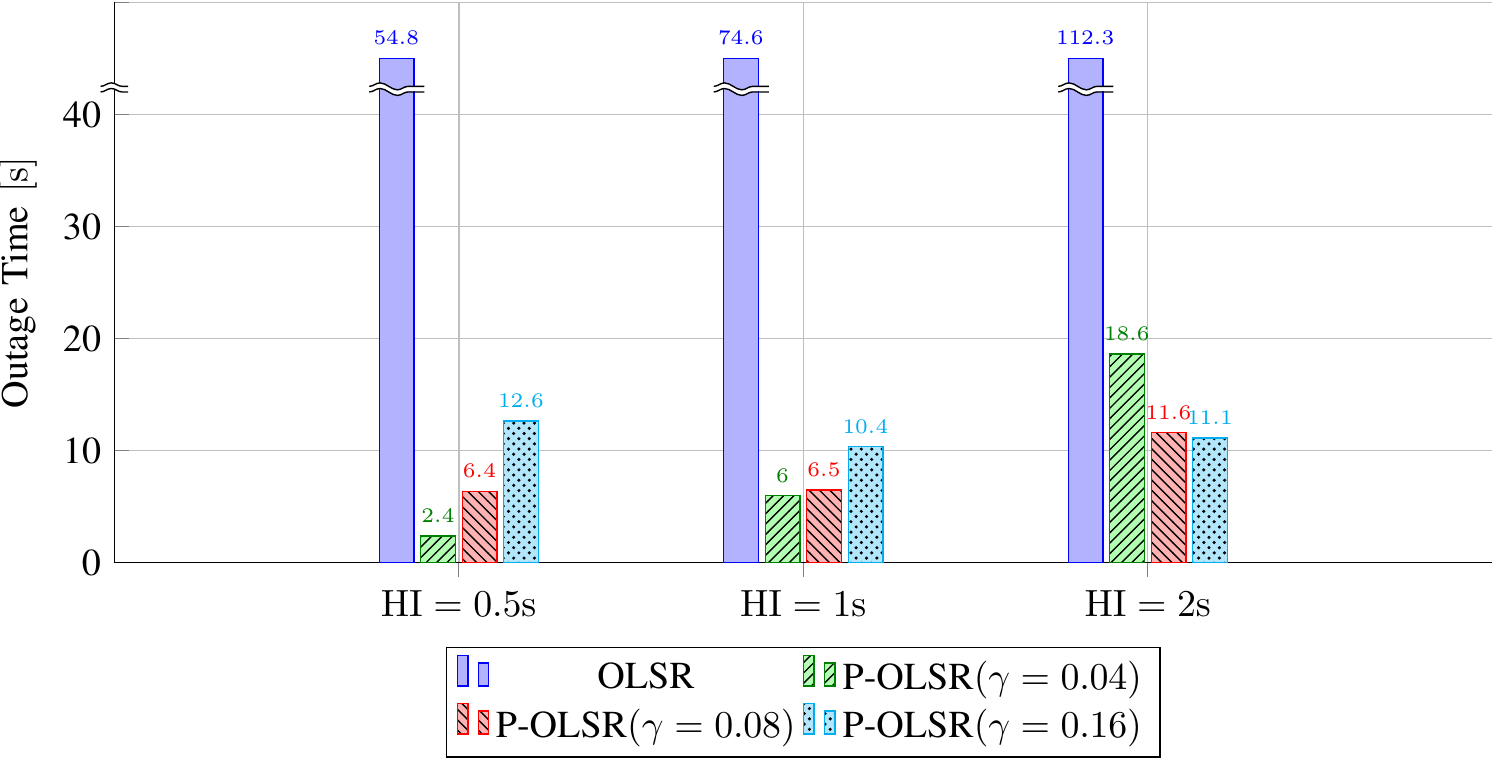}
\caption{Average outage time for OLSR and P-OLSR with $\alpha=0.2$, $\beta=0.2$, and different $\mathrm{HI}$ values.}
\label{fig:outage1}
\end{figure}

\begin{figure}
\centering
%\scalebox{0.95} % Change this value to rescale the drawing.
\includegraphics[width=1.0\columnwidth]{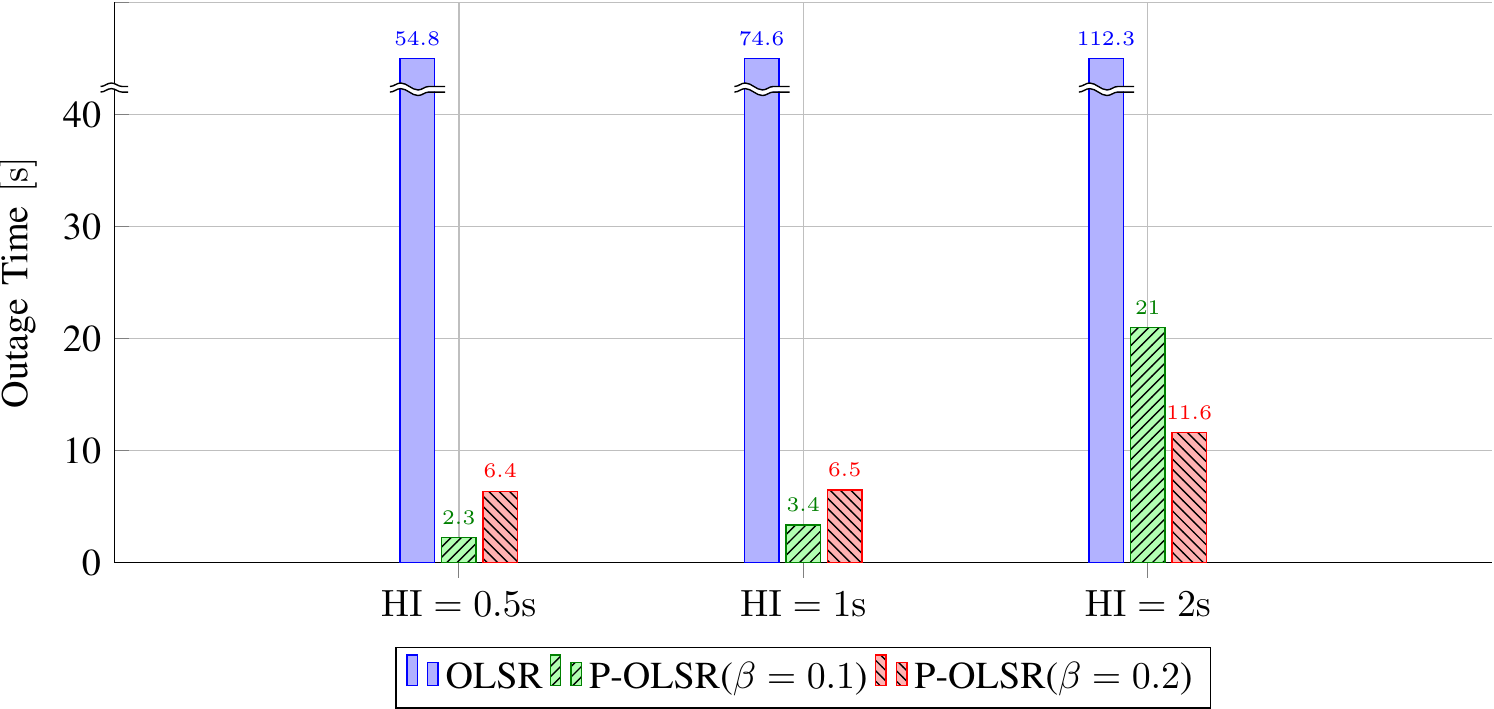}
\caption{Average outage time for OLSR and P-OLSR with  $\alpha=0.2$, and $\gamma=0.08$, and different $\mathrm{HI}$ values. }
\label{fig:outage2}
\end{figure}

In Figs. \ref{fig:outage1} and  \ref{fig:outage2}  we compare OLSR and P-OLSR for different durations of the $\mathrm{HI}$ while keeping the aging parameter  $\alpha$  equal to 0.2.
We present several P-OLSR configurations: In Fig. \ref{fig:outage1}, we maintain a fixed value of $\beta$, ($\beta=0.2$) and we vary the speed aging parameter $\gamma$.
In Fig. \ref{fig:outage2}, we fix $\gamma$   ($\gamma=0.08$) and we vary the value of $\beta$.
Notice that P-OLSR reduces drastically the outage time compared to OLSR. In all its configurations, P-OLSR cuts down the outage time by at least 85\%. 
Considering only the best performing P-OLSR configurations, the outage reduction is about 95\%, 92\%, and 90\%, for $\mathrm{HI}$ durations equal to 0.5, 1, and 2 seconds, respectively.

In Fig. \ref{fig:outage1}, we  compare P-OLSR configurations for different values of $\gamma$. 
We see that when the $\mathrm{HI}$ is short, it is convenient to decrease the value of $\gamma$, and vice versa. The reason is that if the {\it Hello} message rate is low, it is convenient to have a fast aging  of the previous speed estimates. 
In Fig. \ref{fig:outage2}, we  examine the role of the parameter $\beta$. We notice that when the $\mathrm{HI}$ is long, it is more convenient to increase  $\beta$, meaning  that  the speed term has a greater  weight in the EXT computation, (see Eq. \eqref{etx2}). 
Intuitively, when the {\it Hello} message rate is low, the link-quality estimation is slow in tracking the channel fluctuations, therefore it is convenient to give more weight to the speed term 
 than to the link-quality term. 

\begin{figure}
\centering
%\scalebox{0.95} % Change this value to rescale the drawing.
\includegraphics[width=1.0\columnwidth]{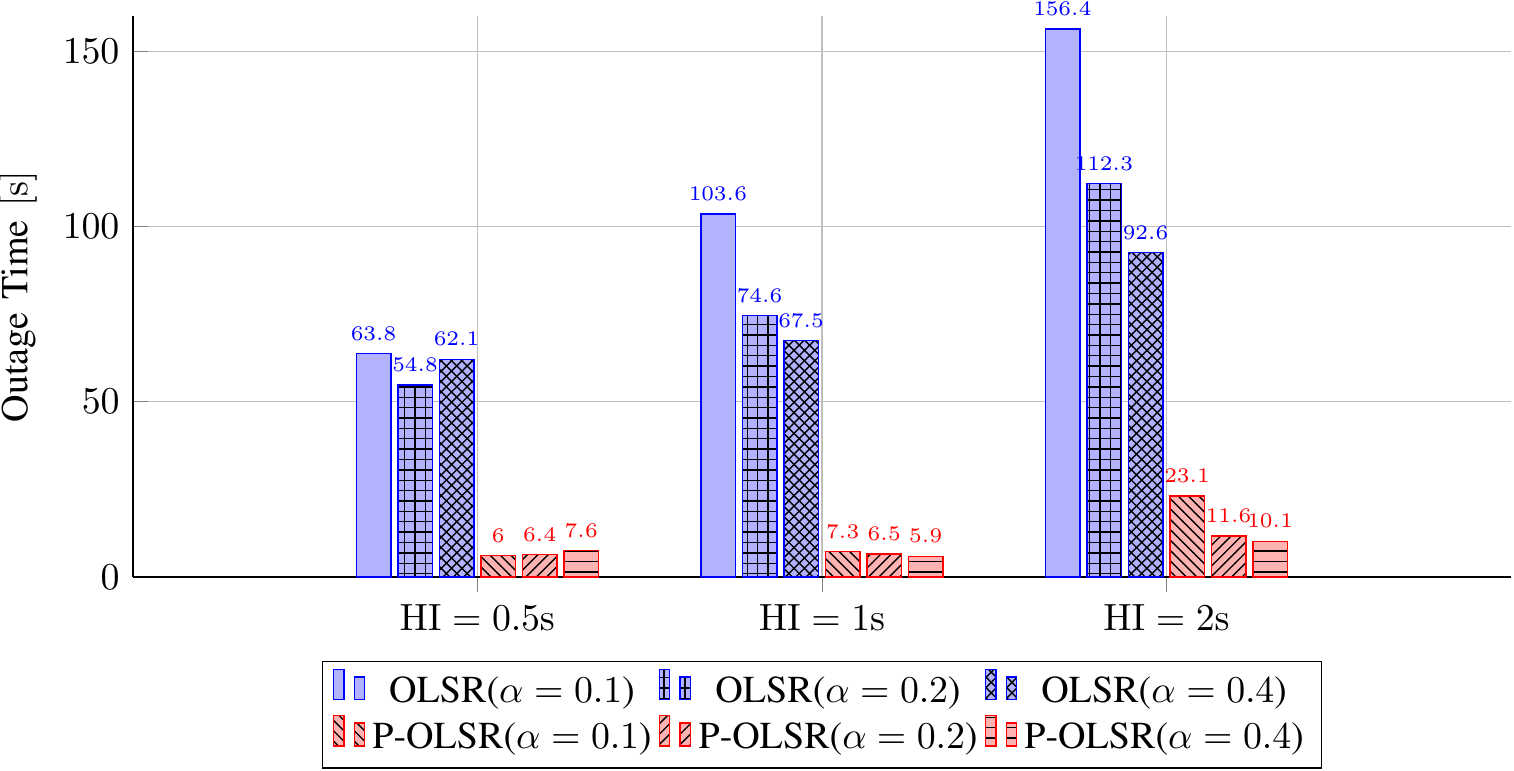}
\caption{Average outage time for OLSR and P-OLSR with different $\mathrm{HI}$ values and $\alpha$. }
\label{fig:outage3}
\end{figure}

In Fig. \ref{fig:outage3}, we can examine the role of the  $\mathrm{HI}$ and the link quality  aging, $\alpha$.  
We compare P-OLSR and OLSR considering different values of aging and different $\mathrm{HI}$ durations. To avoid overcrowding the plots, we consider only one P-OLSR configuration, that is ($\beta=0.2$ and $\gamma=0.08$).

As we notice in Fig. \ref{fig:outage3}, the shorter the $\mathrm{HI}$ the better the performance. 
However the improvement brought by halving the $\mathrm{HI}$ from 1 to  0.5 seconds might not be  large enough to justify  the increase of the signalling traffic.  
This is more evident when P-OLSR is used. 
In fact, by exploiting the GPS information, P-OLSR is able to track the evolution of the network topology also when the {\it Hello} message rate is lower.  
As an  example, we note that P-OLSR with $\mathrm{HI}$ of 2 seconds reduces the outage time by 80\% with respect to OLSR with $\mathrm{HI}=0.5$ seconds, even if it generates about 1/4 of the {\it Hello} messages.

\begin{figure}
 \centering
% %\scalebox{0.95} % Change this value to rescale the drawing.
\includegraphics[width=1.0\columnwidth]{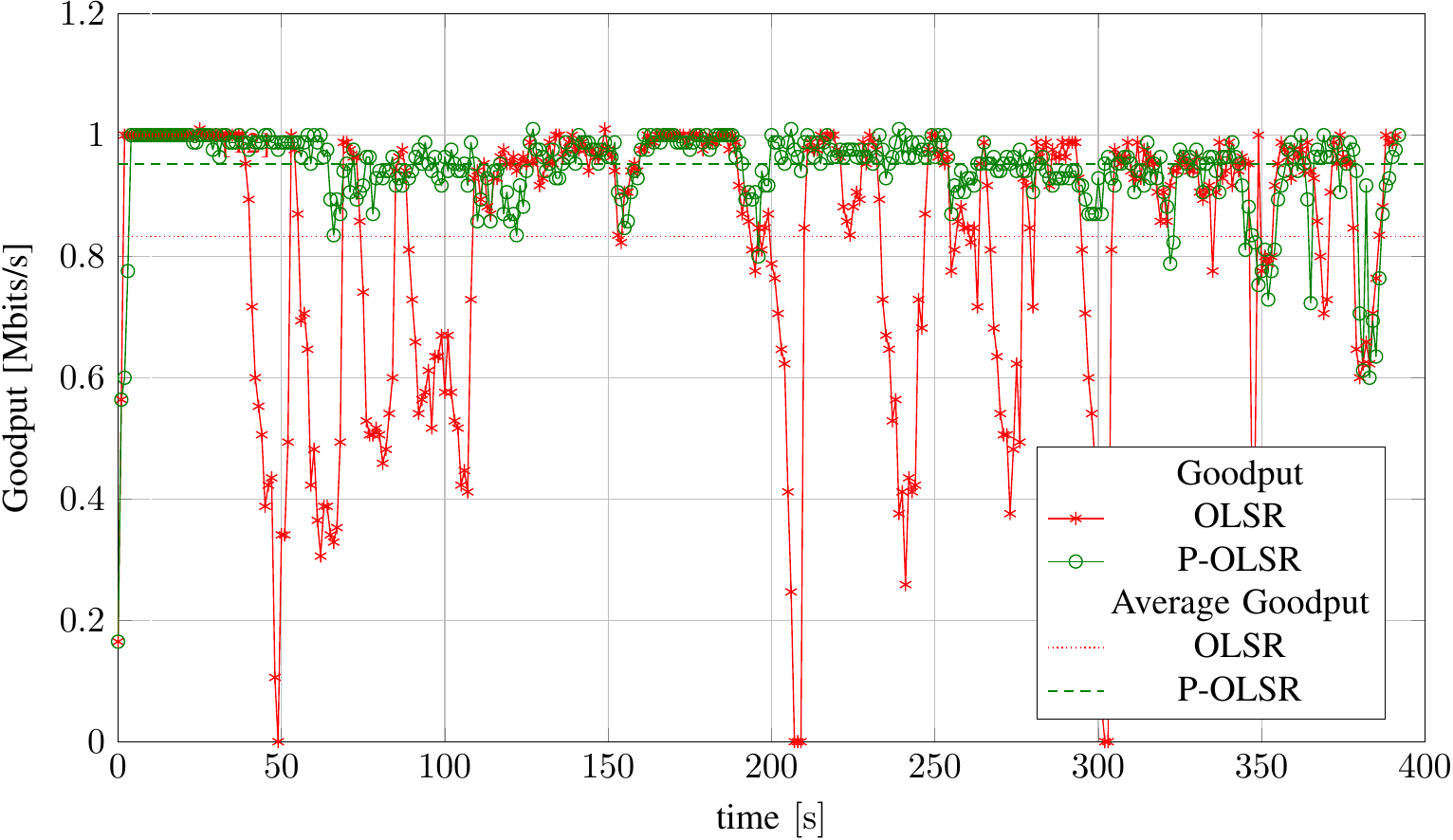}
 \caption{Goodput during an emulation, considering $\mathrm{HI}=1$ second,  $\alpha=0.1$, $\beta=0.2$, and  $\gamma=0.08$.}
 \label{fig:goodput} 
 \end{figure}

\begin{figure}
\centering
%\scalebox{0.95} % Change this value to rescale the drawing.
\includegraphics[width=1.0\columnwidth]{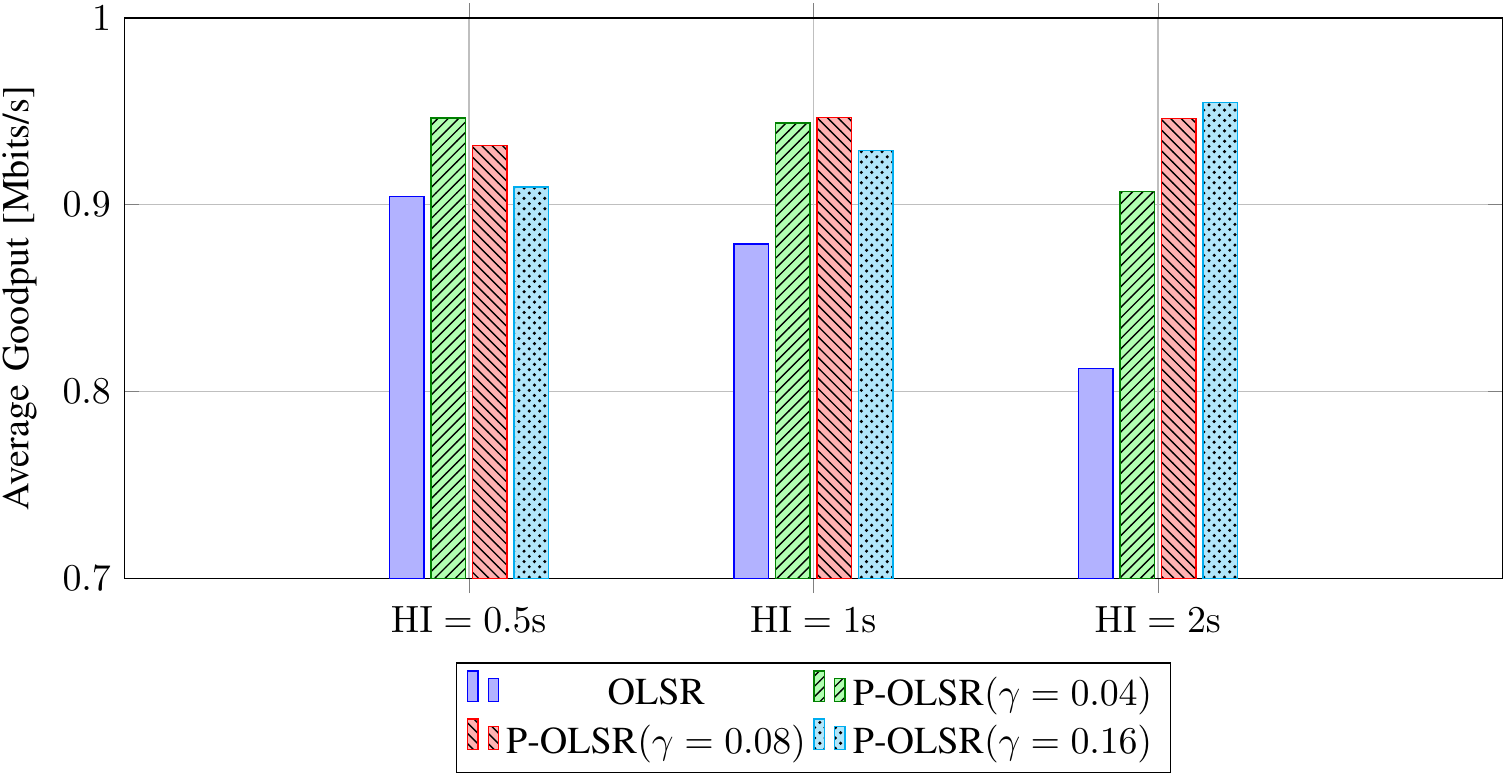}
\caption{Average goodput for OLSR and P-OLSR with $\alpha=0.2$, $\beta=0.2$, and different $\mathrm{HI}$ values. }
\label{fig:bandwidth1}
\end{figure}

\begin{figure}
\centering
%\scalebox{0.95} % Change this value to rescale the drawing.
\includegraphics[width=1.0\columnwidth]{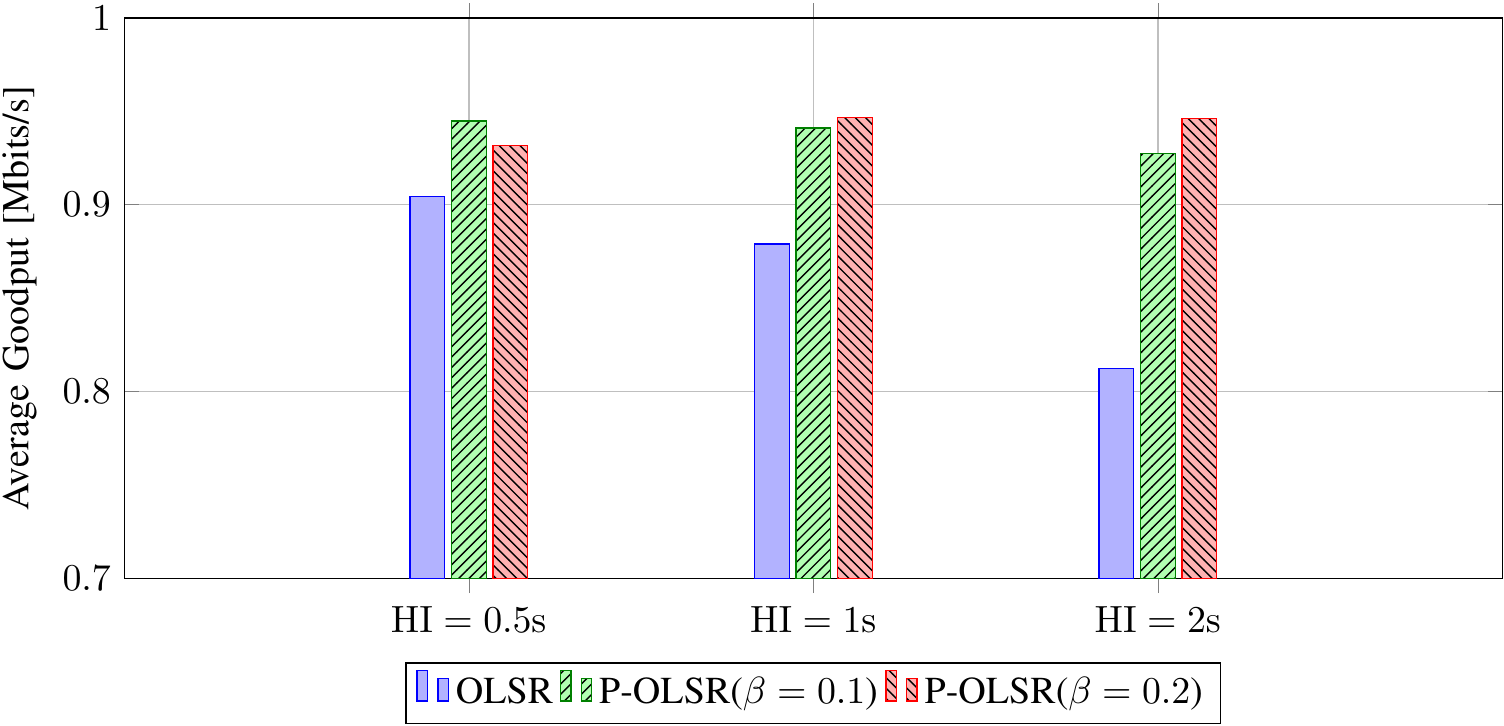}
\caption{Average goodput for OLSR and P-OLSR with  $\alpha=0.2$, and $\gamma=0.08$, and different $\mathrm{HI}$ values. }
\label{fig:bandwidth2}
\end{figure}

\begin{figure}
\centering
%\scalebox{0.95} % Change this value to rescale the drawing.
\includegraphics[width=1.0\columnwidth]{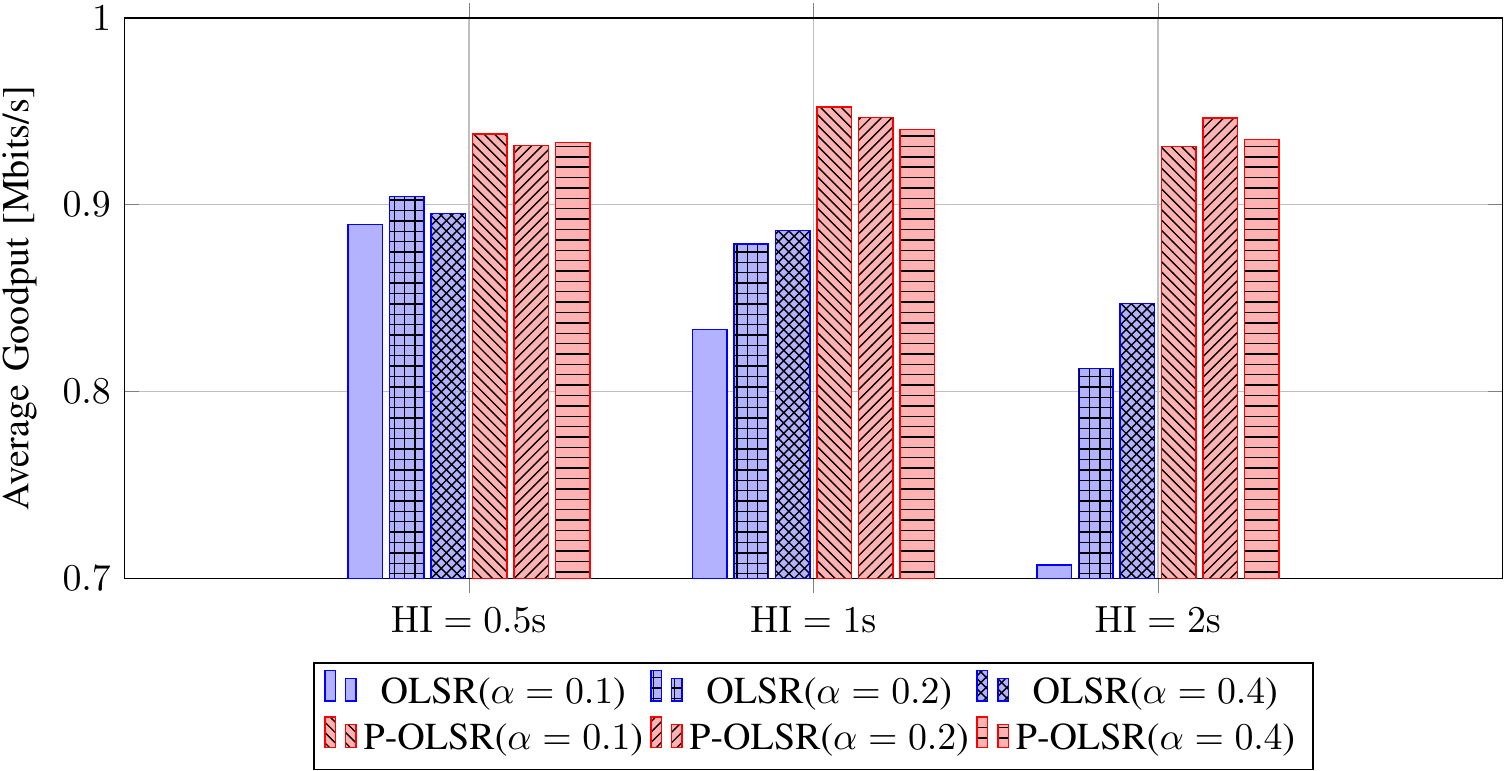}
\caption{Average goodput for OLSR and P-OLSR with different $\mathrm{HI}$ values and $\alpha$. }
\label{fig:bandwidth3}
\end{figure}

We also compare the performance in terms of goodput. In perfect link conditions, the goodput is equal to 1 Mbit/s.
In Fig. \ref{fig:goodput}, we plot the goodput achieved by OLSR and P-OLSR during an emulation with  $\mathrm{HI}=1$ second, $\alpha=0.1$, and the specific-P-OLSR parameters $\gamma=0.08$, $\beta=0.2$.
As we can see from the plot, the goodput achieved by OLSR is unstable. It often drops below 0.6 Mbit/s, and it also reaches zero. This is because OLSR does not reacts promptly to the topology changes.
On the other hand, P-OLSR, which reacts promptly to topology changes and avoids outages, achieves a more stable goodput. It never drops under 0.6 Mbit/s and most of the time it is above 0.8 Mbit/s. The goodput is lower than 1 Mbit/s because some packets are lost due to random channels fluctuations. 
In the plot, we report also  the average goodput over the whole emulation duration, which is 0.83 Mbit/s for OLSR and 0.95 Mbit/s for P-OLSR.
All the other cases behave similarly.
For the sake of  more compact representations, for the other cases we report the average goodput over the emulation duration (Figs. \ref{fig:bandwidth1}-\ref{fig:bandwidth3}).
As before,  we repeated the emulation 10 times for each configuration and report the average results.
In Figs. \ref{fig:bandwidth1} and  \ref{fig:bandwidth2}, we compare OLSR and P-OLSR for several $\mathrm{HI}$ values while keeping $\alpha=0.2$.
In Fig. \ref{fig:bandwidth1}, we fix the value of $\beta$, ($\beta=0.2$) and vary $\gamma$. In Fig. \ref{fig:bandwidth2}, we fix the value of $\gamma$ ($\gamma=0.08$) and vary $\beta$.
We see that in all cases, P-OLSR increases the average goodput compared to OLSR. 
The gap is greater when the {\it Hello} message rate is lower ($\mathrm{HI}=2$ seconds), because OSLR is slower to reacts to topology changes. 
In Fig. \ref{fig:bandwidth3}, we compare P-OLSR and OLSR considering different values of aging and different $\mathrm{HI}$ durations. Again, to avoid overcrowding the plots, we consider only the P-OLSR configuration with $\beta=0.2$ and $\gamma=0.08$.  
Once again, P-OLSR outperforms OLSR in every configuration. The gap is smaller when the {\it Hello} message rate is high.

\section{Conclusion}
\label{conclusions}

In this paper we compared the performance of P-OLSR and that of OLSR in a FANET composed of small fixed-wing UAVs.

Such networks are characterized by a high degree of mobility, which constitutes a challenge to the routing protocol.
Routing protocols designed for MANETs mostly fail in tracking the evolution of the network topology. 
We address this problem by designing an OLSR extension, called P-OLSR: it takes advantage of the GPS information to predict how the quality of the wireless links will evolve. 
The network emulations and fields experiments confirm our expectations.
With P-OLSR, the routing  follows the topology changes without interruptions, which is not the case with OLSR. 
We make use of P-OLSR  to improve the routing in the SMAVNET II project \cite{smavnet_website_download}. 
SMAVNET II is a research project for developing a self-organizing UAV network. 

\section*{Acknowledgments} This work is supported by armasuisse, competence sector Science+Technology for the Swiss Federal Department of Defense, Civil Protection and Sport.

\bibliographystyle{../../bibliography/IEEEtran} % citation style
% the file  bibliography/smavnet.bib is the bibliography database for this document, add entries to it as necessary
\bibliography{../../bibliography/smavnet,../../bibliography/IEEEfull}
\addcontentsline{toc}{part}{\small References}

\begin{IEEEbiography}[{\includegraphics[width=1in,height=1.25in,clip,keepaspectratio]{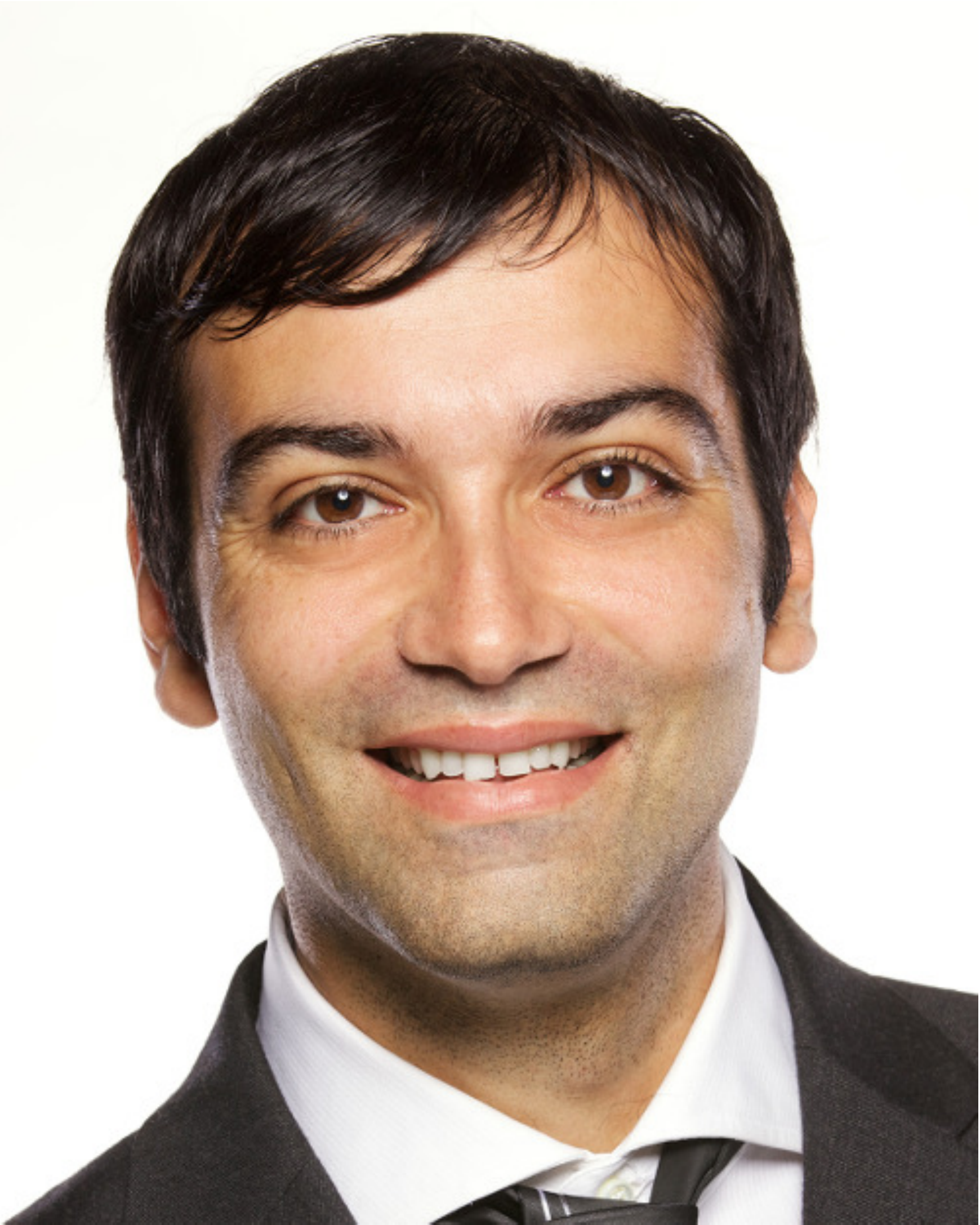}}]{Stefano Rosati}
(S'07-M'11)
is a Post-Doc Research Engineer at the \'Ecole Polytechnique F\'ed\'erale de Lausanne (EPFL), Switzerland. 
He received the Laurea degree (summa cum laude) and Ph.D. degree in Telecommunications Engineering from the University of Bologna, Italy, in 2007 and 2011, respectively.
From 2007 to 2011, he was with  the Advanced Research Center for Electronic Systems (ARCES) of the University of Bologna. 
In 2010, he was an Intern Engineer at Corporate R\&D Department of Qualcomm Inc. (San Diego, CA).
In 2011 he joined the Information Processing Group (IPG) and the Mobile Communications Laboratory (LCM) at  EPFL.

His interests are in various aspects of digital communications, in particular next-generation cellular communication systems, and both terrestrial and satellite broadcast networks.
His research activities are also focused on flying ad-hoc networks and self-organizing networks of Unmanned Aerial Vehicles (UAVs).
He has authored several scientific papers and internationals patents regarding these topics.

\end{IEEEbiography}

\begin{IEEEbiography}[{\includegraphics[width=1in,height=1.25in,clip,keepaspectratio]{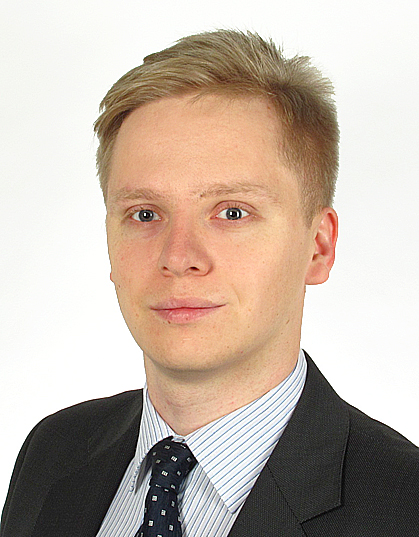}}]{Karol Kru\.zelecki}(M'09) 
is a research engineer working at EPFL, Lausanne, Switzerland.
He received his Msc. Eng. degree in Computer Science and in
Computational Mechanics from Cracow University of Technology, Poland,
in September and November 2008 respectively.
Between 2008 and 2011 he was working at The European Organization for
Nuclear Research (CERN) on the development of automatic build and test
system used to improve the software quality and reliability for the
Large Hadron Collider beauty experiment (LHCb). In 2012 he joined the
Information Processing Group (IPG) and the Mobile Communications
Laboratory (LCM) of the EPFL.
\end{IEEEbiography}

\begin{IEEEbiography}[{\includegraphics[width=1in,height=1.25in,clip,keepaspectratio]{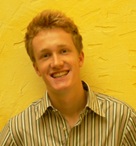}}]{Gr\'egoire Heitz}
is an engineer working at the Laboratory of Intelligent System at EPFL since September 2012. He obtained a M.Sc in Electronic from the ENSCPE of Lyon (France) in 2012. He was working in senseFly SA for his master thesis, working on the development of a new flying robot in 2011. His research interest lies in embedded systems development.
\end{IEEEbiography}

\begin{IEEEbiography}[{\includegraphics[width=1in,height=1.25in,clip,keepaspectratio]{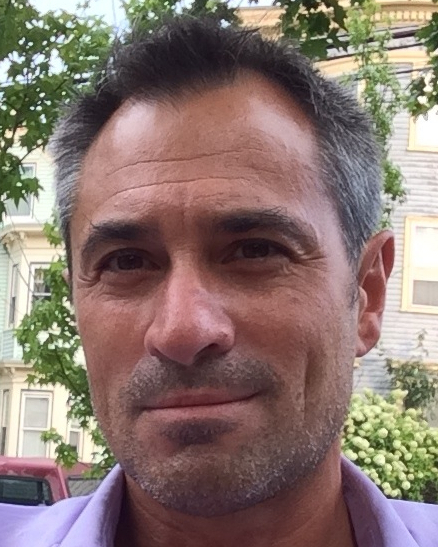}}]{Dario Floreano}
(SM'06) received the M.A. and Ph.D. degrees from the University of Trieste, Trieste, Italy, in 1988 and 1995, respectively, and the M.S. degree from the University of Stirling, Stirling, Scotland, in 1991. He is Full Professor at the \'Ecole Polytechnique F\'ed\'erale de Lausanne, Lausanne, Switzerland, where he is Director of the Laboratory of Intelligent Systems and Director of the Swiss National Center of Competence in Robotics. His research interests are at the convergence of biology, artificial intelligence, and robotics. He authored more than 300 peer-reviewed articles and 3 books on the topics of evolutionary robotics, bio-inspired artificial intelligence, and biomimetic flying robots, and spun two companies off
\end{IEEEbiography}

\begin{IEEEbiography}[{\includegraphics[width=1in,height=1.25in,clip,keepaspectratio]{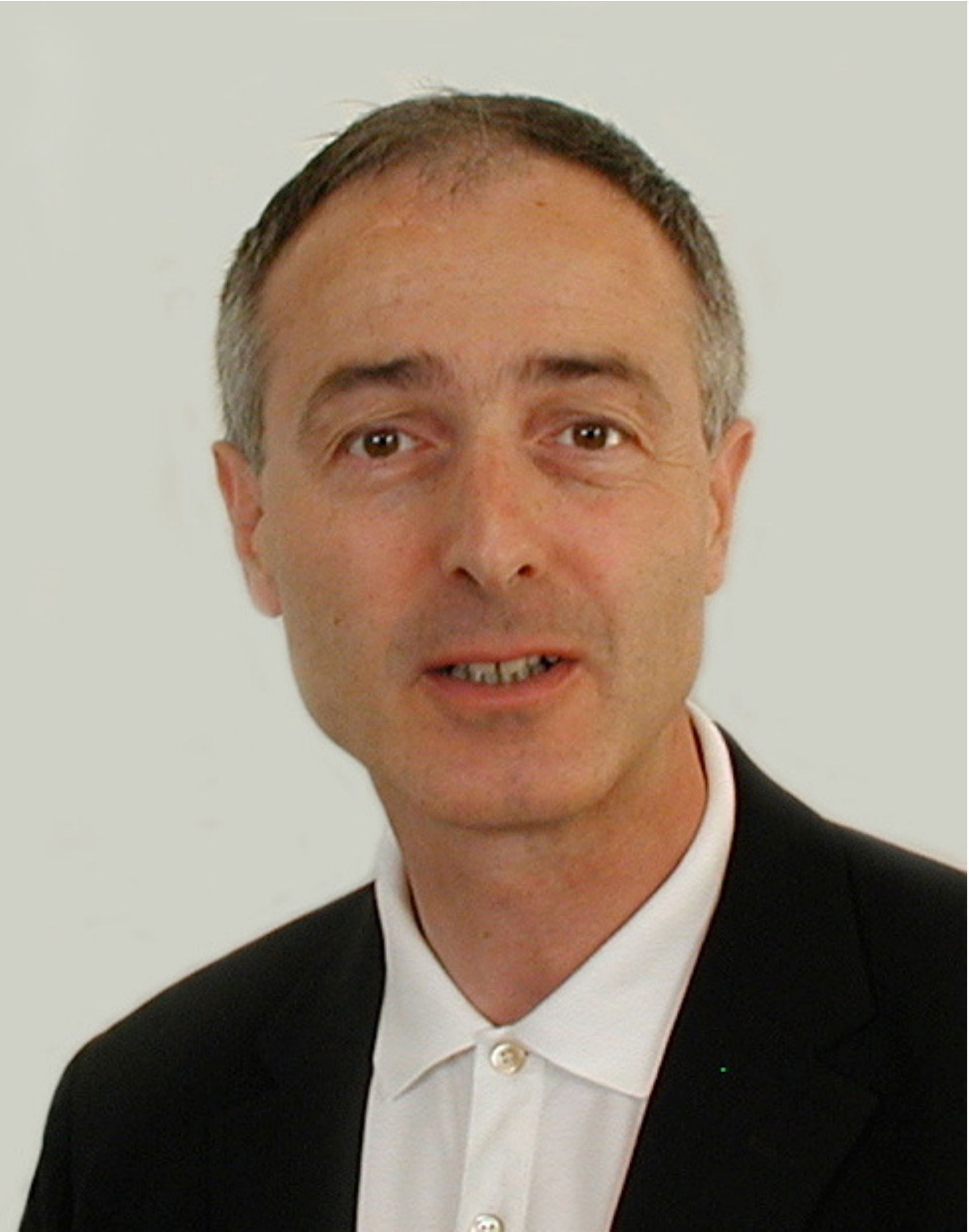}}]{Bixio Rimoldi} (S'83-M'85-SM'92-F'00) received his Diploma and his Doctorate from the Electrical Engineering department of the Eidgeno\"{o}ssische Technische Hochschule in Zurich (ETHZ). During 1988-1989 he held visiting positions at the University of Notre Dame and Stanford. In 1989 he joined the faculty of Electrical Engineering at Washington University, St. Louis, and since 1997 he is a Full Professor at the \'Ecole Polytechnique F\'ed\'erale de Lausanne (EPFL) and director of the Mobile Communications Lab. He has spent sabbatical leaves at MIT (2006) and at the University of California, Berkeley (2003-2004). 

In 1993 he received a US National Science Foundation Young Investigator Award. In 2000 he was elected to the grade of Fellow of the IEEE. During the period 2002-2009 he has been on the Board of Governors of the IEEE Information Theory Society where he served in several offices including President. He was co-chairman with Bruce Hajek of the 1995 IEEE Information Theory Workshop on Information Theory, Multiple Access, And Queueing (St Louis, MO), and co-chairman with Jim Massey of the 2002 IEEE International Symposium in Information Theory (Lausanne, Switzerland). He was a member of the editorial board of "Foundations and Trends on Communications and Information Theory," and was an editor of the European Transactions on Telecommunications. During 2005 and 2006 he was the director of EPFL's undergraduate program in Communication Systems. 

His interests are in various aspects of digital communications, in particular information theory, and software-defined radio.
\end{IEEEbiography}

\end{document}